\documentclass[10pt, final, journal, letterpaper,twocolumn]{IEEEtran}
\makeatletter
\def\ps@headings{%
\def\@oddhead{\mbox{}\scriptsize\rightmark \hfil \thepage}%
\def\@evenhead{\scriptsize\thepage \hfil \leftmark\mbox{}}%
\def\@oddfoot{}%
\def\@evenfoot{}}
\makeatother 
\UseRawInputEncoding
\IEEEoverridecommandlockouts
\usepackage{bbm}
\usepackage{amsfonts}
\usepackage[dvips]{graphicx}
\usepackage{times}
\usepackage{cite}
\usepackage{amsmath}
\usepackage{array}
\usepackage{amssymb}

\usepackage{stfloats}
\usepackage{slashbox}
\usepackage{graphicx}
\usepackage{footnote}
\usepackage{booktabs}
\usepackage{array}
\usepackage{algorithm}
\usepackage{subeqnarray}
\usepackage{cases}
\usepackage{threeparttable}
\usepackage{color}
\usepackage{hyperref}
\usepackage{epstopdf}
\usepackage{algpseudocode}
\usepackage{bm}
\usepackage{multirow}
\usepackage[labelformat=simple]{subcaption}
\usepackage{changepage}
\usepackage{adjustbox}

\begin{document}
\title{Deep Learning for Wireless Coded Caching with Unknown and Time-Variant Content Popularity}

\author{\authorblockN{Zhe Zhang and Meixia Tao, \emph{Fellow, IEEE}}
\thanks{This work is supported by the National Key R\&D Project of China under grant 2019YFB1802702.
Part of this work was presented in IEEE/CIC ICCC 2019 \cite{ICCC}. (\emph{Corresponding author: Meixia Tao})
The authors are with Dept. of Electronic Engineering, Shanghai Jiao Tong University, Shanghai, China. Emails: \{zhang\_zhe, mxtao\}@sjtu.edu.cn. Code is available at: https://github.com/GroupTao/DL-for-Wireless-Coded-Caching.}
}

\maketitle

%
%

\begin{abstract}
Coded caching is effective in leveraging the accumulated storage size in wireless networks by distributing different coded segments of each file in multiple cache nodes.
This paper aims to find a wireless coded caching policy to minimize the total discounted network cost, which involves both transmission delay and cache replacement cost, using tools from deep learning.
The problem is known to be challenging due to the unknown, time-variant content popularity as well as the continuous, high-dimensional action space.
We first propose a clustering based long short-term memory (C-LTSM) approach to predict the number of content requests using historical request information. This approach exploits the correlation of the historical request information between different files through clustering. Based on the predicted results, we then propose a supervised deep deterministic policy gradient (SDDPG) approach. This approach, on one hand, can learn the caching policy in continuous action space by using the actor-critic architecture.
On the other hand, it accelerates the learning process by pre-training the actor network based on the solution of an approximate problem that minimizes the per-slot cost.
Real-world trace-based numerical results show that the proposed prediction and caching policy using deep learning outperform the considered existing methods.
\end{abstract}
\begin{IEEEkeywords}
Coded caching, clustering, LSTM, deep reinforcement learning, supervised learning.
\end{IEEEkeywords}
\section{Introduction}
\subsection{Background and Contribution}
Recent years have witnessed rapid development in rich media-enabled
applications on mobile devices, such as YouTube and Youku.
The plenty of data requests and large data size result in considerable traffic burden in both core networks and wireless access networks.
A promising solution, known as edge caching, is to store the popular contents at the edge of wireless networks during off-peak periods when the network resources are abundant. During peak-traffic times, the cached contents can be served immediately upon user requests without being fetched from the core network.
In this manner, edge caching can effectively alleviate the traffic burden, reduce the transmission delay, and improve the user experience. It thus has received tremendous attention from both academia and industry. Existing caching techniques can be roughly divided into two categories, uncoded caching and coded caching. In uncoded caching, each file is either cached entirely without partitioning or not cached at all in a cache node.  In coded caching, each file can be partitioned into multiple segments, which are then encoded using e.g., maximum distance separable (MDS) code and cached distributively in different nodes. In general, coded caching outperforms uncoded caching as the former can better utilize the accumulated cache size among different nodes. A brief overview of the seminal works and the recent development on caching techniques can be found in \cite{JSAC18}.

Exploiting the promises of edge caching depends highly on the knowledge of the content popularity, which, however, is often unknown in advance. The content popularity may also change dynamically over time and space due to the arrival of new contents and the user mobility in wireless environment.
To this end, machine learning has emerged as a powerful tool to predict the content popularity or user request based on historical observations, and then make the cache decision to optimize certain objective functions in wireless networks.

Despite the tremendous previous works on this topic, learning-based caching still remains as a challenging problem due to the following factors. First, cache-equipped base stations or helper nodes in wireless networks usually can only access the user request information rather than user or file context. Therefore, traditional context-aware prediction algorithms used for recommendation systems cannot be applied.  Second, due to the limited coverage range, each base station can only communicate with a limited number of users. As such, compared with cloud-based learning, the dataset available in each base station is very sparse and hence it is difficult to predict user request accurately. Third, the popularity prediction and cache placement strategy are tightly coupled because the cache decision will affect the accuracy of popularity prediction, and in turn, the predicted popularity will affect the cache decision. Last but not least, in a multi-cell network, the caching decision among multiple nodes should be made collaboratively and therefore suffers from the curse of dimensionality.

In this work, we address some of the above challenges by applying deep learning techniques for wireless coded caching in multi-cell networks when the content popularity is unknown and time-variant. Similar to the network model in \cite{femto}, each file is first encoded by an MDS code, then stored distributively at different cache nodes. A user can recover its requested file as long as it collects sufficient number of coded bits of the file from its nearby cache nodes or a macro base station. The cached contents at each cache node will be updated in response to the latest predicted content requests. Our contribution is two-fold. We first propose a clustering-based long short-term memory (C-LSTM) prediction framework by using clustering and recurrent neural network (RNN) to online predict the number of content requests. This framework takes the historical request patterns as content features for clustering. The clustering process leverages the correlation in historical request patterns among all files. The clustering results are then sent to a bank of LSTM networks to predict the content requests where each LSTM network corresponds to a cluster. The LSTM network is a widely used RNN which has been proved to be
very effective to address sequence prediction problems such as those found in natural
language processing. Compared with the traditional model-based prediction, the LSTM network is to learn the correlation between sequences without assuming a prediction model in advance. File request patterns are not completely unrelated, and the request patterns for files belonging to the same topical subject are often similar. In addition, due to people’s daily work, file request patterns are usually periodic. As such, the C-LSTM prediction framework is well suited for our problem.

Our second contribution is to propose a supervised deep deterministic policy gradient (SDDPG) approach to learn how much coded fraction of each file should be stored in each cache node based on the predicted content requests. Our objective is to minimize the total discounted network cost that involves both transmission delay and cache replacement cost. 
We formulate this problem as a non-stationary Markov decision process (MDP), for which the deep deterministic policy gradient (DDGP) approach is well-suited as both the system state and action spaces are continuous and of high dimension.
In order to ensure that the output of the actor network in DDPG meets the cache capacity constraint in each cache node, we use the \emph{sigmoid} function as the activation function of the output layer and add a linear scaling after the output layer.
To accelerate the learning process, we use supervised learning to pre-train both the actor and critic networks where the training samples are generated using the solution of an approximate problem that minimizes the per-slot cost instead of the total network cost.

Numerical results based on a real-world dataset show that the proposed C-LSTM achieves higher prediction accuracy than the considered existing methods and the proposed SDDPG approach provides lower total network cost than existing policies. Results also verify that coded caching outperforms uncoded caching in practical environment with unknown content popularity.

\subsection{Related Works}
Machine learning has been widely used for content popularity estimation and cache strategy learning in wireless networks.  
In \cite{zipfTCOM,transfer2}, a transfer learning approach is proposed to improve the estimation of popularity profile by leveraging prior information obtained from a surrogate domain.
In \cite{Contcaching}, the authors take the diversity in content popularity across the users into account and learn the cache strategy through the feedback from environment based on deep deterministic policy gradient algorithm (DDPG).
In \cite{living}, the fixed global content popularity is estimated based on collaborative filtering and then exploited for cache decision to maximize the average user request satisfaction ratio in small-cell networks.
The work \cite{ARAL} considers the minimization of energy cost for systematic traffic transmission under a framework consisting of mobile edge caching and cache-enabled D2D communications.
In \cite{letter}, the authors propose a scheme based on LSTM and external memory to enhance the
decision making ability of the base station.
In \cite{TVT}, a deep reinforcement learning based joint proactive cache placement and power allocation strategy is proposed where a set of nodes cooperatively serve the content request.
Note that the works \cite{Contcaching, living,transfer2,zipfTCOM,ARAL,letter,TVT} assume that the content popularity is time-invariant and hence may not be applicable in practical systems where the user preferences are dynamic.

For time-variant content popularity, there are some works take into account the temporal variation of user preference in cache placement.
In \cite{TWCMAB}, the authors take context information of users into account and use contextual multi-armed bandit (MAB) to learn the context-specific content popularity online.
In \cite{DynamicLocation}, the authors use a linear prediction model to estimate future content hits by leveraging content feature and location differentiation.
In \cite{trendcache}, the authors learn the relationship between the future popularity of contents and their context.
In \cite{DRLA}, two potential recurrent neural networks (RNNs) are adopted to predict user mobility and content popularity based on the context of users.
However, the users' context and content feature information exploited in \cite{TWCMAB, trendcache, DynamicLocation, DRLA} are often unavailable if the cache node is operated by mobile network operators, which in general can only observe local content requests. In our work, we only use the historical content requests which are easy to observe in the practical system.
In \cite{ZNF}, the authors propose a grouped linear regression model to estimate future requests by using historical user requests only and then apply reinforcement learning (RL) to learn the cache strategy.
Therein, the prediction of content requests at each time slot is classified into different groups according to the age of each content, i.e., the time elapsed since the release of the content. Compared with \cite{ZNF}, this work utilizes the correlation between file request patterns, which is more useful as shall be verified with numerical results.
The works \cite{LSTM11,LSTM15,LSTM16,lstm_gru} apply the LSTM network  for content popularity prediction without any model assumption.
Compared to the independent prediction of each time series in \cite{LSTM11,LSTM15,LSTM16,lstm_gru}, clustering considered in our work also allows similar file request patterns to be processed using the same LSTM network which improves prediction accuracy. Another line of work to tackle the time variation of content popularity is to bypass the prediction stage and to directly learn the caching strategy using reinforcement learning as in \cite{JSAC2020} and \cite{Xu}.

The optimization of coded caching with unknown content popularity has been studied in \cite{pmab} and \cite{mugen}. In particular,
the work \cite{pmab} use MAB to learn content popularity modeled by a Zipf distribution and then optimize the coded cache placement in small-cell networks.
In \cite{mugen}, the authors propose a deep reinforcement learning based approach to maximize the successful transmission probability in coded caching enabled fog radio access networks. Note that, both \cite{pmab} and \cite{mugen} focus only on the transient cache decision, not the long-term caching policy.
Different from \cite{pmab} and \cite{mugen} which consider only the impact of the caching strategy on the current moment, we take the cache replacement cost into account and aim to minimize the total discounted weighted-sum of transmission delay and replacement cost of the network over an infinite time horizon.

It is worth noting that there have been a number of works that apply deep neural network (DNN) for various communication tasks. For example, in works \cite{sun,weiyu,yu16}, DNNs are used to solve resource allocation problems. Note that the main purpose of these works \cite{sun,weiyu,yu16} is to reduce the computational complexity of solving the interested optimization problems by utilizing DNN. The use of deep learning techniques in this work is, however, not only to reduce the computational complexity but also to improve the performance of the final result.
\subsection{Organization and Notations }
The rest of the paper is organized as follows. Section II
establishes the system model. Section III proposes a clustering and deep neural network based online prediction approach for the prediction of request number. Section IV proposes a deep reinforcement learning based approach for coded cache placement for total discounted network cost minimization. The performance of the proposed method is provided in Section V. Conclusions are drawn in Section VI.

Boldface lower-case and upper-case letters
denote vectors and matrices respectively. Calligraphy letters
denotes sets. $\mathbb{E}(\cdot)$ denotes the expectation of a random variable.
We use $I(x)$ to denote the indicator function for feature $x$; its value indicates the cluster the feature $x$ belongs to. $\sigma(\cdot)$ denotes the sigmoid function.

\section{Problem Description}
\subsection{Network Model}
\begin{figure}
\begin{centering}
\includegraphics[width=0.46\textwidth]{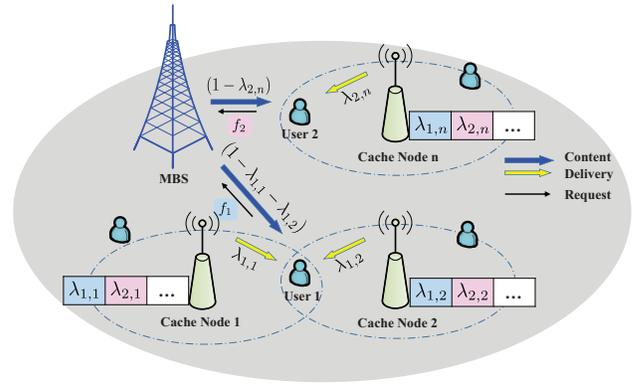}
\caption{{A content delivery network with coded caching.}}\label{fig:systemmodel}
\end{centering}
\vspace{-0.2cm}
\end{figure}

As illustrated in Fig.~\ref{fig:systemmodel}, we consider a wireless caching network with one macro base station (MBS), $N$ cache nodes and $K$ users.
Let $\mathcal{N}^+ \triangleq \{0, 1, \ldots, N\}$ denote the set of cache nodes where the index $n = 0$ represents the MBS.
Let $\mathcal{N}\triangleq \{1, 2,\ldots, N\}$ denote the set of cache nodes only.
Let $\mathcal{K}\triangleq \{1,2,\ldots,K\}$ denote the set of users, where each user can represent a group of users in the same area.
Each cache node has a certain communication range. We let $\mathcal{N}_k \subseteq \mathcal{N}^+$ represent the set of cache nodes (including MBS) to which user $k$ can communicate with. The delay of transmitting one bit from cache node $n$ to user $k$ is denoted as $\delta_{k,n}$, for $n\in \mathcal{N}_k$ and $k\in \mathcal{K}$, and it mainly depends on the communication distance. We sort the cache nodes in each $\mathcal{N}_k$ in the ascending order of the per-bit delay to user $k$, such that $(j)_k$ denotes the index of the cache node with the $j$th shortest delay to user $k$. We assume all users in the system can download files from the MBS but with a much longer per-bit delay, i.e., $\delta_{k,0} > \delta_{k,n}, \forall n\in \mathcal{N}_k\setminus\{0\}$.

We consider a content catalogue consisting of $F$ files, denoted as set $\mathcal{F}=\{1,2,\ldots,F\}$. Each file is assumed to have the same length of $B$ bits. Each cache node can only store up to $M\cdot B$ bits $(M < F)$ and the MBS can access all files in the catalogue. Note that although the content catalogue is modeled as static in this paper, the arrival of new files from content providers in practical systems can be well captured by the time-variant content request distribution as described below. In specific, we consider a time-slotted system.
At each time slot $t\in\{1,2,\ldots \}$, let $\mathbf{d}_k(t) = [d_{k,1}(t),d_{k,2}(t),\ldots, d_{k,F}(t)]$ denote the demand vector of user $k$, where $d_{k,f}(t)\geq0$ represents the number of requests for file $f \in \mathcal{F}$. Considering the dynamic arrival of new contents, if file $f$ is not available yet at time slot $t$, then $d_{k,f}(t)=0, \forall k\in \mathcal{K}$.
Further let $\mathbf{d}(t) = [\mathbf{d}_1(t),\mathbf{d}_2(t),\ldots, \mathbf{d}_K(t)]$ denote the demand vector of all users at time slot $t$.
We assume that the duration of each time slot $t$ is long enough so that all user requests can be served within a slot.

Similar to \cite{femto}, we assume that each file is encoded by a rateless MDS code and the coded bits are stored independently and distributively at the cache nodes. With MDS coding, a file can be retrieved provided that $B$ coded bits are collected in any order from the MBS or the cache nodes.
Let $\bm{\Lambda}_n(t) = [\lambda_{1,n}(t),\lambda_{2,n}(t),\ldots, \lambda_{F,n}(t)]$ denote the cache vector variable of cache node $n\in\mathcal{N}$ at time slot $t$, where $\lambda_{f,n}(t)\in [0,1]$ represents the fraction of coded bits from file $f \in \mathcal{F}$ cached at this node.
Further let $\bm{\Lambda}(t) = [\bm{\Lambda}_1(t),\bm{\Lambda}_2(t),\ldots, \bm{\Lambda}_N(t)]$ denote the cache vector of all cache nodes.
A more practical case where each $\lambda_{f,n}$ can only take values from a finite and discrete set shall be considered in the simulation.
The delay of downloading a fraction of coded bits $\lambda_{f,n}(t)B$ on the link from cache node $n\in \mathcal{N}_k$ to user $k\in \mathcal{K}$ is given by $\lambda_{f,n}(t) \delta_{k,n}B$.
If user $k$ can retrieve its requested file $f$ at time slot $t$ from the coded bits collectively stored by its best $j$ cache nodes, the delay is given by
\begin{equation}
D_k^{f,j}(t)=B\sum_{i=1}^{j-1}\lambda_{f,(i)_k}(t)\delta_{k,(i)_k}+B\left(1-\sum_{i=1}^{j-1}\lambda_{f,(i)_k}(t)\right)\delta_{k,(j)_k}.
\end{equation}
Note that the complete file $f$ can be downloaded by user $k$ from its best $j$ cache nodes only if $\sum_{i=1}^{j-1}{\lambda_{f,(i)_k} < 1}$ and $\sum_{i=1}^j{\lambda_{f,(i)_k}\geq 1}$. In the special case where $j=|\mathcal{N}_k|$, user $k$ downloads the uncached fraction from the MBS, i.e., $(j)_k=0$. For example, in Fig.~\ref{fig:systemmodel}, user 1 requests $f_1$ and receives the fraction $\lambda_{1,1}$ and $\lambda_{1,2}$ of $f_1$ from the cache node 1 and 2 respectively. The remaining fraction $(1-\lambda_{1,1}-\lambda_{1,2})$ is obtained from the MBS. According to \cite[Lemma 6]{femto}, the delay for user $k$ to download file $f$ is
\begin{equation}
D_k^{f}(t)=\max_{j\in\{1,2,\ldots,|\mathcal{N}_k|\}}D_k^{f,j}(t).
\end{equation}
Hence, the total transmission delay to meet all user requests $\mathbf{d}(t)$ in time slot $t$ is given by
\begin{equation}
C_\text{d}(t)=\sum_{k\in\mathcal{K}}\sum_{f\in\mathcal{F}}d_{k,f}(t) D_k^{f}(t).
\end{equation}

In this paper, we also consider the replacement cost for updating cached content. We define the replacement cost at each time slot as the total increment of the cached file fraction as compared with the previous time slot over all cache nodes and files. Specifically, the replacement cost can be expressed as
\begin{equation}
C_\text{r}(t)=\sum_{f\in\mathcal{F}}\sum_{n\in\mathcal{N}}\max\{\lambda_{f,n}(t)-\lambda_{f,n}(t-1),0\}.
\end{equation}
Considering both the transmission delay and the replacement cost, we define the network cost at time slot $t$ as
\begin{equation}
C(t)=C_\text{d}(t)+\beta C_\text{r}(t),
\end{equation}
where $\beta \geqslant 0$ is a weighting factor to balance the two costs.
%
\subsection{Problem Formulation}
We aim to minimize the expectation of the total discounted network cost over an infinite time horizon by optimizing the coded cache
placement. The problem can be formulated as

\begin{subequations}
\begin{align}
\mathcal{P}(t):&\min\limits_{\{\bm{\Lambda}(t)\}} \lim\limits_{T\rightarrow\infty}\mathbb{E}\left [ \sum_{t=1}^T\gamma^{t-1}(C_\text{d}(t)+\beta C_\text{r}(t))\right ]\\
\label{equ:frac} &\hspace{+0.14cm}\text{s.t.}\hspace{+0.31 cm} \sum_{f\in \mathcal{F}}\lambda_{f,n}(t)\leq M, \forall n\in\mathcal{N}, \forall t\\
\label{equ:size} &\hspace{+1.0 cm}\lambda_{f,n}(t)\in[0,1], \forall f\in\mathcal{F}, n\in\mathcal{N}, \forall t,
\end{align}
\end{subequations}
where $\gamma \in [0,1]$ is a discount factor which reflects the impact of future network cost on current cache decision. Since the instantaneous content requests $\mathbf{d}(t)$ at each time slot cannot be foreseen before making the caching decision
$\bm{\Lambda}(t)$, the problem is intractable.

\begin{figure}[t]
\begin{centering}
\includegraphics[width=0.46\textwidth]{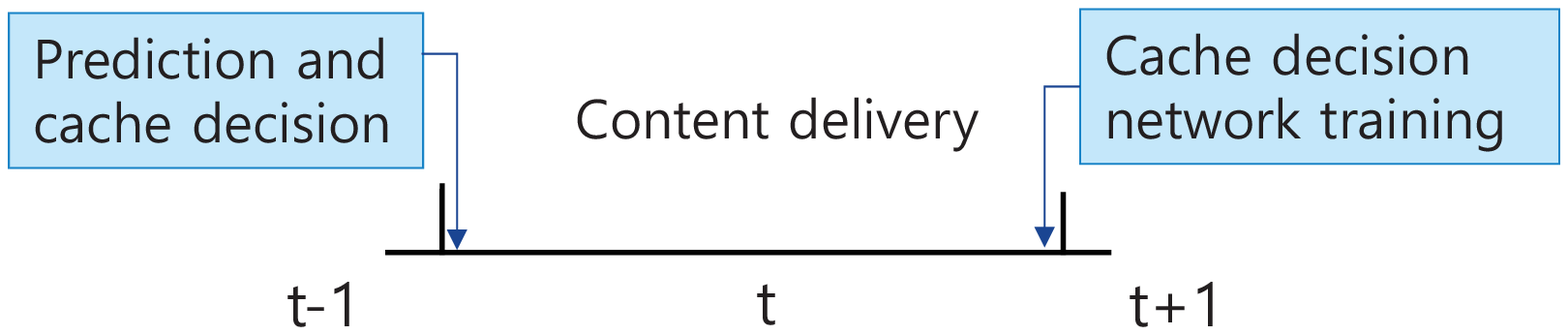}
\caption{The sequential diagram of the system operation.}\label{fig:time}
\end{centering}
\vspace{-0.2cm}
\end{figure}

In the next two sections, we shall introduce the proposed C-LSTM approach for request prediction and SDDPG approach for cache decision to solve (6). The overall execution sequence of the proposed approach in each time slot is shown in Fig.~\ref{fig:time}.
At the beginning of each time slot, we first predict the number of requests for contents recorded in the catalogue according to historical information.
Based on the predicted number of requests and the cache status of the previous time slot, the cache decision network decides the cache allocation of the current time slot. 
During the time slot, users submit file requests and the content delivery phase occurs.
After content delivery, we update the parameters of prediction model according to the actual requests. In the meantime,
the cache decision network is trained based on the actual network cost observed from the environment.

\section{Request Prediction Using Clustering and LSTM}
The traditional prediction methods assume that the files are independent, so each file corresponds to a prediction network. However, due to the large number of files in the core network, this will consume a lot of computing resources. In fact, due to the daily work of people, most of the file request patterns are periodic, such as one day or one week. In addition, files with similar context are usually requested by the same person, so their received request patterns are similar.
In this section, we first classify the time-variant content requests using K-means clustering by exploiting the above-mentioned correlation in historical request patterns among all files. We then utilize the LSTM network to predict the content requests at each time slot for each cluster. We refer to the overall algorithm as the clustering-LSTM (C-LSTM) prediction method.


\subsection{K-means Clustering for Content Requests}

Since the number of requests per user is usually small and hence difficult to predict, we predict the total number of requests per file in each time slot instead. Let $\widetilde{\mathbf{d}}(t) =[\widetilde{d}_{1}(t), \widetilde{d}_{2}(t), \ldots, \widetilde{d}_{F}(t)]$ where $\widetilde{d}_{f}(t)$ denotes the predicted number of requests for file $f \in \mathcal{F}$ from all users at time slot $t$, i.e., the estimation of $d_f(t) = \sum_{k\in \mathcal{K}} {d_{k,f}(t)}$.

Let $\mathbf{p}_{t,f}\in\mathbb{R}^{\rho}$ denote a $\rho$-dimensional feature vector of file $f$ at time slot $t$, which is defined as:
\begin{equation}
\mathbf{p}_{t,f}=[d_f(t-\rho),d_f(t-\rho+1),\ldots,d_f(t-1)].
\end{equation}
The feature vector $\mathbf{p}_{t,f}$ contains the historical requests of file $f$ during the previous $\rho$ time slots, where $ \rho$ is a design parameter. To leverage the correlation in historical request patterns among all files for content request prediction, we propose to partition the observed feature vectors of all files into $C$ clusters, where $C$ is a design parameter. By clustering, the request prediction of a file can not only use the request information of this file but also the request information from other files and hence can be more accurate.
In addition, since the number of clusters $C$ is usually much less than the number of files $F$, the number of samples that each cluster can use to train the prediction network is greatly increased compared to the independent prediction of each file.

We adopt the K-means \cite{kmean} clustering algorithm for classification.
In order to eliminate the difference in the absolute value of the request number, let $\bar{\mathbf{p}}_{t,f}=\{\bar{d}_f(t-\rho),\bar{d}_f(t-\rho+1),\ldots,\bar{d}_f(t-1)\}$ denote the normalized feature vector, where $\bar{d}_f(t-\rho)=d_f(t-\rho)/\max \{ \mathbf{p}_{t,f}\}$ and $\max \{ \mathbf{p}_{t,f}\}$ presents the maximum value of all elements in $\mathbf{p}_{t,f}$. The similarity between two feature vectors is characterized by the Euclidean distance between their normalized counterparts. To begin with (i.e. $t=\rho+1$), we adopt the method in \cite{kmean++} to select $C$ points from the $F$ feature vectors $\{\bar{\mathbf{p}}_{\rho+1,1},\bar{\mathbf{p}}_{\rho+1,2},\ldots,\bar{\mathbf{p}}_{\rho+1,F}\} $, denoted as $\{ \mathbf{p}_1^c,\mathbf{p}_2^c,\ldots,\mathbf{p}_{C}^c\}$, to be the initial cluster centers.
Specifically, we first randomly select one of the $F$ feature vectors as the first initial cluster center $\mathbf{p}_1^c$. We then select each subsequent initial cluster center at random with a probability proportional to the distance from itself to the closest center that has been already chosen.

At the beginning of each time slot $t\geq \rho+1$, we determine the cluster  membership of the newly-observed feature vectors $\bar{\mathbf{p}}_{t,f}, \forall f\in \mathcal{F}$, according to the minimum Euclidean distance criterion. More specifically:
\begin{equation}
I(\bar{\mathbf{p}}_{t,f}) = \arg \min\limits_{i\in{1,2,\ldots,C}}
||\bar{\mathbf{p}}_{t,f}-\mathbf{p}_i^c||_2^2.
\end{equation}

At the end of time slot $t$, we update the center of each cluster by averaging the feature vectors within this cluster. In specific, the new center of cluster $i$ will be

\begin{equation}
\mathbf{p}_i^c=\frac{\mathbf{p}_i^c S_i(t-1)+\sum_{I(\bar{\mathbf{p}}_{t,f})=i,f\in\mathcal{F}} \bar{\mathbf{p}}_{t,f}}
{S_i(t-1)+\sum_{I(\bar{\mathbf{p}}_{t,f})=i,f\in\mathcal{F}} 1 },
\end{equation}
where $S_{i}(t)$ represents the accumulated number of feature vectors in the cluster $i$ at the end of time slot $t$. With the increase of time slot $t$, the center of each cluster gradually becomes stable. As a result, the correlation between feature vectors belonging to the same cluster is getting stronger and stronger. The correlation reflects the similarity in the trend of changes in the number of file requests.
\subsection{LSTM Network for Request Prediction}
The LSTM network \cite{LSTM1} is a widely used recurrent neural network (RNN) for processing sequential data and has been found extremely successful in many applications, such as speech recognition \cite{ls1}, machine translation \cite{ls2}, parsing \cite{ls3} and image captioning \cite{ls4}. Therefore, it is well suited for our considered content request prediction problem. In this subsection, we first briefly introduce the structure of the LSTM network.
Then, based on the results of clustering in the previous subsection, we propose a cluster-specific LSTM-based prediction framework that can update the network parameters online so as to gradually improve the accuracy of prediction.

\begin{figure}[t]
\begin{centering}
\includegraphics[width=0.45\textwidth]{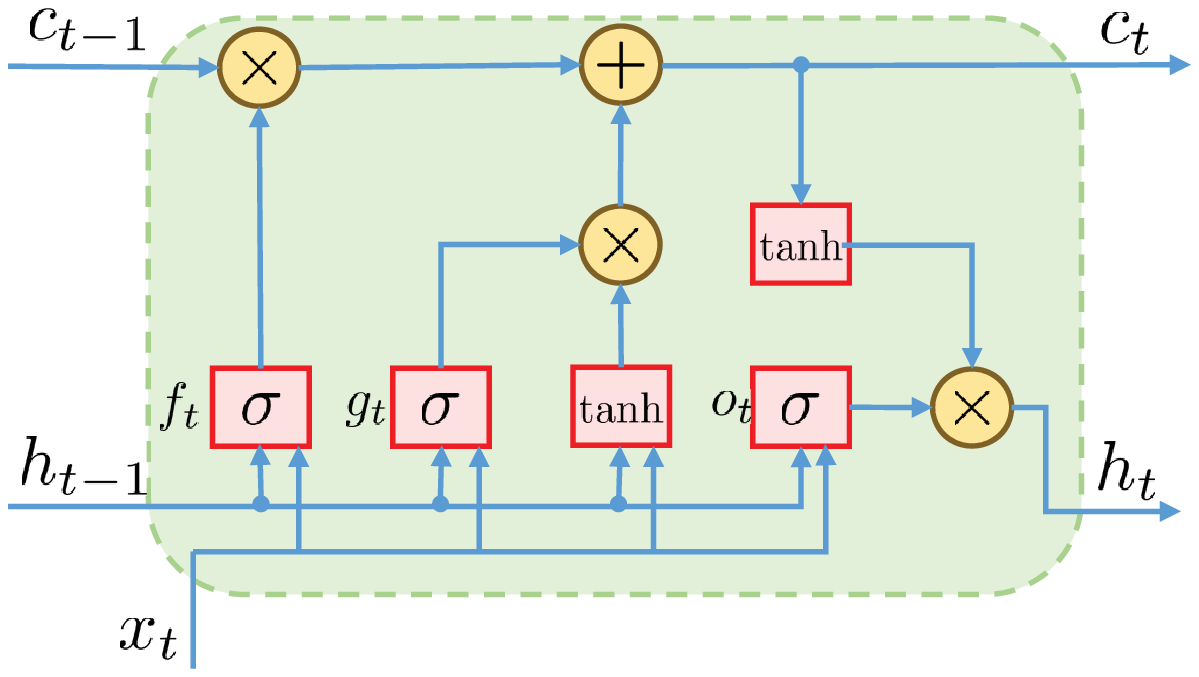}
 \caption{{The structure of LSTM cell.}}\label{fig:LSTM unit}
\end{centering}
\vspace{-0.3cm}
\end{figure}

The LSTM network is composed of multiple copies of basic memory blocks and each memory block contains a memory cell. The block diagram of LSTM cell is shown in Fig.~\ref{fig:LSTM unit}. The input gate $g_t$, forget gate $f_t$ and output gate $o_t$ are all sigmoid units to optionally pass information. As its name implies, $f_t$ and $g_t$ respectively decide the forgetting amount of the internal state $c_{t-1}$ and the updating amount of the new one. The output $h_t$ of LSTM cell can also be shut off, via the output gate $o_t$.
LSTM cells are connected recurrently to each other. At each moment $t$, LSTM cell updates the state $c_t$ and generates corresponding output $h_t$ (the predicted content request $\widetilde{d}_f(t)$) according to the cell state $c_{t-1}$ and output $h_{t-1}$ at the
previous moment, as well as the input of the current moment $x_t$ (the normalized content request $\bar{d}_f(t-1)$), and pass them to the next moment.

\begin{figure}[t]
\begin{centering}
\includegraphics[width=0.47\textwidth]{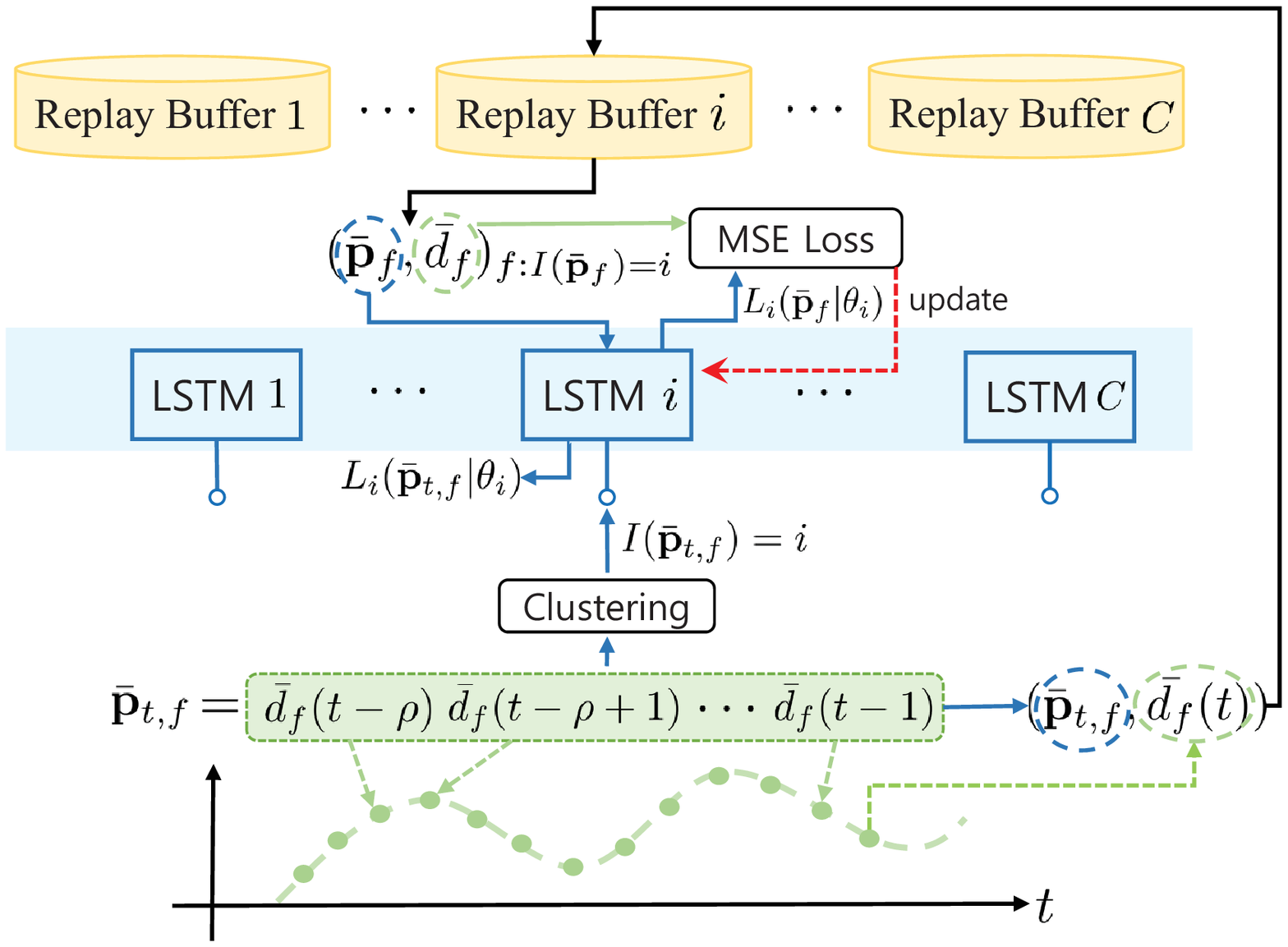}
 \caption{{The overall diagram of LSTM prediction.}}\label{fig:LSTM_prediction}
\end{centering}
\end{figure}

For each cluster, we use an LSTM network with three hidden LSTM-layers,
one input layer, and one output layer to do the prediction.
Let $L_i(\cdot)$ denote the output function of LSTM network corresponding to cluster $i$, parameterized by a set $\theta_i$.
At the beginning of time slot $t$, we first determine the cluster membership of the
newly-observed feature vectors $\bar{\mathbf{p}}_{t,f}$ according to the method proposed in the previous subsection. For example, if $I(\bar{\mathbf{p}}_{t,f})=i$, the predicted number of request $\widetilde{d}_f(t)$ is given by
\begin{equation}
\widetilde{d}_f(t)=\max\{\mathbf{p}_{t,f}\}L_i(\bar{\mathbf{p}}_{t,f}|\theta_i).\label{equ:pre}
\end{equation}
In order to improve the stability of training, we use the replay buffer to record training samples. Let $\mathcal{R}_i^L$ denote the replay buffer of size $S_i^L$ corresponding to cluster $i$.
At the end of time slot $t$, the normalized feature vectors and the actual number of requests are added as training samples to the replay buffer of the corresponding cluster. For example, if $I(\bar{\mathbf{p}}_{t,f})=i$, add $(\bar{\mathbf{p}}_{t,f},d_f(t)/\max\{\mathbf{p}_{t,f}\})$ to the replay buffer $\mathcal{R}_i^L$. The oldest sample will be discarded when the replay buffer is full.
Then, the LSTM network is updated by sampling a minibatch uniformly from the corresponding replay buffer. The $i$-th LSTM network is updated by minimizing the loss between the predicted number of request and the actual number of request, defined as:
\begin{equation}
Loss(\theta_i)=(L_i(\bar{\mathbf{p}}_{t,f}|\theta_i)-d_f(t)/\max\{\mathbf{p}_{t,f}\})^2. \label{equ:lstm loss}
\end{equation}
The overall diagram of LSTM prediction is shown in Fig.~\ref{fig:LSTM_prediction}. With the increase of time slot $t$, the normalized feature vectors stored in the same replay buffer become more and more similar which makes it easier for LSTM network to learn the request pattern represented by each cluster.

Overall, the proposed C-LSTM prediction algorithm is outlined in Alg.~\ref{predict}

\begin{algorithm}[t]
\caption{Clustering-LSTM Prediction for Content Requests}\label{predict}
\begin{algorithmic}[1]
\State Initialize $C$ cluster centers.
\State Randomly initialize each LSTM network $L_i(\bar{\mathbf{p}}|\theta_i)$ with weights $\theta_i, \forall i \in \{1,2,\ldots,C\}$.
\State Initialize each replay buffer $\mathcal{R}_i^L$ of size $S_i^L, \forall i \in \{1,2,\ldots,C\}$.
    \For{$t= \rho+1:T$}
    \State Phase 1: Request Prediction at the beginning of time slot $t$. 
    \For{$f= 1:F$}
    \State Generate and normalize the feature vectors $\mathbf{p}_{t,f}$ according to historical requests.
    \State Cluster $\bar{\mathbf{p}}_{t,f}$ according to the Euclidean distance from $\bar{\mathbf{p}}_{t,f}$ to each cluster center.
    \State Use LSTM network to predict the number of requests according to (\ref{equ:pre}).
    \EndFor
    \State Phase 2: Network Training at the end of time slot $t$. 
    \For{$i= 1:C$}
    \State Update the center point and the size of cluster $i$.
    \State Store new samples  $\{(\bar{\mathbf{p}}_{t,f},d_f(t)/\max\{\mathbf{p}_{t,f}\})|I(\bar{\mathbf{p}}_{t,f})=i\}$ in the replay buffer $\mathcal{R}_i^L$.
    \State Get a random minibatch of $M_L$ samples from replay buffer $\mathcal{R}_i^L$.
    \State Train the LSTM network $L_i$ by minimizing the loss function (\ref{equ:lstm loss}).
    \EndFor
    \EndFor
\end{algorithmic}
\end{algorithm}


\section{Cache Decision Using Supervised DDPG}
In this section, we introduce the proposed SDDPG approach to learn the coded caching policy based on the results of request prediction. The SDDPG approach accelerates the learning process of the existing DDPG algorithm with supervised learning. In the following, we first introduce the deep reinforcement learning framework, then introduce supervised learning to pre-train the neural network according to the solution of an approximate problem that minimizes per-slot network cost.


\subsection{Deep Reinforcement Learning Framework}
Problem (6) can be viewed as a real-time control problem which can be solved with RL. The essential elements of RL, i.e.,  state, action, reward and return are defined as follows:
\begin{itemize}
  \item \textbf{State}: The state of system at time slot $t$ is defined as:
      \begin{equation}
      {s}_t=[\mathbf{\widetilde{d}}(t),\bm{\Lambda}(t-1)],
      \end{equation}
      where $\mathbf{\widetilde{d}}(t)$ is the predicted number of requests for files at time slot $t$ and $\bm{\Lambda}(t-1)$ is the cache status in the previous time slot.
  \item \textbf{Action}: The action at time slot $t$, denoted as ${a}_t$, is defined as the cache allocation $\bm{\Lambda}(t)$ which represents the fraction of coded bits of each file that should be cached in each cache node.
  \item \textbf{Reward}: The reward at time slot $t$ is defined as the negative actual network cost $-C(t)$ observed at the end of time slot $t$, denoted as $r_t(s_t,a_t)$.
  \item \textbf{Return}: The \emph{return} at time slot $t$ is defined as the sum of discounted future reward from $t$:
      \begin{equation}
      R_t = \lim\limits_{T\rightarrow\infty}\sum_{i=t}^T\gamma^{i-t}r_i(s_i,a_i).
      \end{equation}
\end{itemize}

Note that the return also depends on the action, and therefore on the policy.
Let $\mu: \mathcal{S} \to \mathcal{A}$ denote the policy mapping any state $s\in\mathcal{S}$ to any action $a\in\mathcal{A}$. We model the problem (6) as a non-stationary Markov decision process (MDP) with an initial state distribution $p(s_1)$ and transition dynamics $p(s_{t+1}|s_{t},\mu (s_t))$.
Our goal is to learn a policy which maximizes the expected reward from the start distribution $J_\mu = \mathbb{E}_{r_i,s_i\sim E}[R_1]$,
where $E$ stands for the environment.
The action-value function describes the expectation of return $R_t$ after taking an action $a_t$ following policy $\mu$ in state $s_t$. It can be rewritten recursively as
\begin{align}
Q_{\mu}(s_t,a_t)&= \mathbb{E}_{r_{i\geq t},s_{i\geq t} \sim E}[R_t|s_t, \mu(s_t)] \notag \\
                &= \mathbb{E}_{r_t,s_{t+1}\sim E}[r_t(s_t,\mu(s_t))+\gamma (R_{t+1}|s_{t+1},\mu(s_{t+1})] \nonumber \\
                &= \mathbb{E}_{r_t,s_{t+1}\sim E}[r_t(s_t,\mu(s_t))+\gamma Q_{\mu}(s_{t+1},\mu(s_{t+1}))].
\end{align}
\begin{figure}[t]
\begin{centering}
\includegraphics[width=0.50\textwidth]{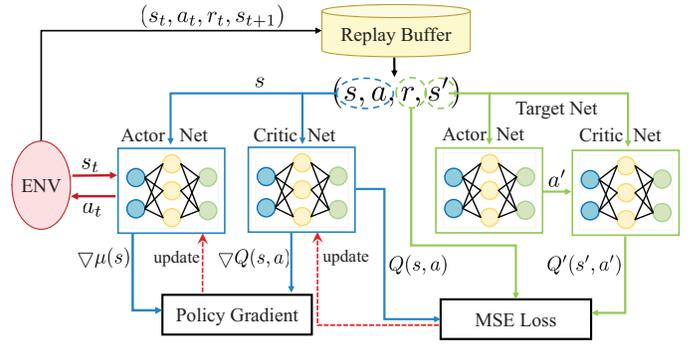}
 \caption{{The overall diagram of DDPG.}}\label{fig:DDPG}
\end{centering}
\vspace{-0.2cm}
\end{figure}
Since both state and action are high-dimensional and continuous, the traditional Q-learning cannot be applied. We adopt the DDPG method \cite{DDPG}, an actor-critic architecture based on the deterministic policy gradient that can operate over continuous and high-dimensional action spaces. The overall diagram of DDPG is shown in Fig.~\ref{fig:DDPG}.
It uses a critic network parameterized by $\theta_Q$ to approximate the action-value function $Q(\cdot)$ and an actor network parameterized by $\theta_\mu$ to approximate the policy $\mu$. The experience replay is to train the networks with minimum correlation while the target network is designed to slowly track the learned network.


Let the replay buffer be denoted as $\mathcal{B}$ of size $S$.
At the beginning of time slot $t$, the actor generates a proto action $\mu(s_t|\theta_\mu)$ based on the current state $s_t$. To overcome the challenge of action exploration in the learning procedure, we add noise sampled from an Ornstein-Uhlenbeck process, denoted as $\alpha$, to the proto action, i.e., $a_t=\mu(s_t|\theta_{\mu})+\alpha$. At the end of time slot $t$, we first aggregate reward $r_t$ from the environment and store the experienced transition tuple $(s_t,a_t,r_t,s_{t+1})$ as a training sample in the replay buffer $\mathcal{B}$. The oldest sample will be discarded when the replay buffer is full. Then, the actor and critic networks are updated by sampling a minibatch uniformly from the buffer.
In particular, the critic network is updated using MSE loss function defined as:
\begin{equation}
  Loss(\theta_Q)=(Q(s_t,a_t|\theta_Q)-y_t)^2,
\end{equation}
where $y_t=r(s_t,a_t)+\gamma Q(s_{t+1},\mu(s_{t+1}|\theta_{\mu})|\theta^{Q})$.
The actor network is updated by applying the chain rule to the expected return $J_\mu$ with respect to the actor parameters: 
\begin{align}
      \nabla_{\theta_\mu}J_\mu &\approx \nabla_{\theta_\mu}Q(s,a|\theta_Q)|_{s=s_t,a=\mu(s_t|\theta_\mu)}     \notag \\
                            &=\nabla_{a}Q(s,a|\theta_Q)|_{s=s_t,a=\mu(s_t|\theta_\mu)}\nabla_{\theta_\mu}\mu(s|\theta_{\mu})|_{s=s_t}. \label{equ:a_gradient}
\end{align}
The above is the policy gradient \cite{Dgradient}.

The target network is a copy of the actor and critic networks, $Q'(s,a|\theta_Q^{'})$ and $\mu'(s|\theta_{\mu}^{'})$. It is introduced to improve the stability of learning. The weights of these target networks are updated by having them slowly track the learned networks: $\theta' \leftarrow \tau\theta + (1-\tau)\theta'$ with $\tau \ll 1$. Note that DDPG is an off-policy algorithm, the replay buffer size $S$ should be as large as possible to allow the algorithm to take advantage of a large set of uncorrelated experienced samples. 

\begin{figure}[t]
\begin{centering}
\includegraphics[width=0.45\textwidth]{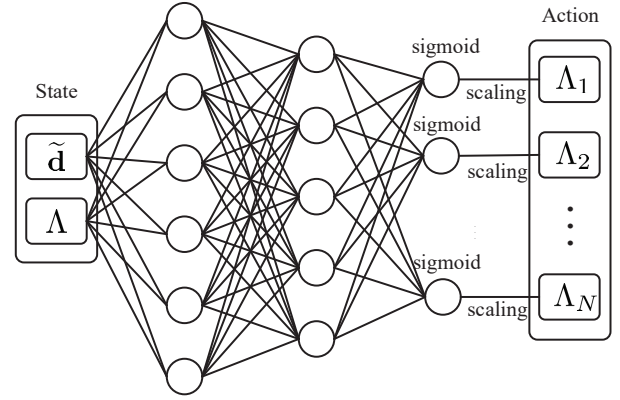}
 \caption{{The structure of actor network.}}\label{fig:actor}
\end{centering}
\vspace{-0.4cm}
\end{figure}

In order to better use DDPG algorithm to solve the cache decision problem, we specially design the structure of the actor network and the critic network.
For the actor network, in order to ensure the actor output meets the cache fraction constraint (\ref{equ:frac}),
we use sigmoid function, $f(x) = 1/(1+ e^{-x})$, as the activation function of the output layer.
Furthermore, to meet the cache capacity constraint (\ref{equ:size}), we introduce a scaling process to each actor output. In specific, the cache vector corresponding to cache node $n$ after scaling is given by
\begin{equation}
\bm{\Lambda}_n(t)=\frac{M}{\sum_{f\in\mathcal{F}}\widetilde{\lambda}_{f,n}(t)}[\widetilde{\lambda}_{1,n}(t),\widetilde{\lambda}_{2,n}(t),\ldots,\widetilde{\lambda}_{F,n}(t)] ,
\end{equation}
where $[\widetilde{\lambda}_{1,n}(t),\widetilde{\lambda}_{2,n}(t),\ldots,\widetilde{\lambda}_{F,n}(t)]$ represents the network output corresponding to cache node $n$ at time slot $t$.
If some elements of cache vector variable are greater than 1 after scaling, we just let them equal 1. Note that when the original output of the actor network before scaling exceeds the cache capability, the elements in the cache vector variable will become smaller after scaling, thus yielding reduction in feedback rewards of the environment.
Since DDPG aims to maximize the discounted future reward, this will, to a certain extent, restrict the actor from generating solutions that exceed the cache capacity. The structure of the actor network is shown in Fig.~\ref{fig:actor}.

\begin{algorithm}[H]
\caption{SDDPG algorithm}\label{SDDPG}
\begin{algorithmic}[1]
\State Randomly initialize critic network $Q(s,a|\theta_Q)$ and actor network $\mu(s|\theta_{\mu})$ with weights $\theta_Q$ and $\theta_{\mu}$.
\State Initialize replay buffer $\mathcal{B}$ and a random process $\alpha$.
\State Initialize target network $Q'$ and $\mu'$ with weights $\theta_Q^{'} \leftarrow \theta_{Q}$, $\theta_{\mu}^{'} \leftarrow \theta_{\mu}$.
\State Phase 1-1: Pre-train the actor network. 
    \For{$t=2:T_A$}
    \State Solve $P_A(t)$ and add $((\mathbf{\widetilde{d}}(t),\bm{\Lambda}(t-1)), \bm{\Lambda}^*(t))$ into $\mathcal{T}_A$.
    \EndFor
    \State Train the actor network by minimizing $Loss(\theta_\mu)$.
    \State Update the target actor network: $\theta_{\mu}^{'}\leftarrow \theta_{\mu}$
\State Phase 1-2: Pre-train the critic network. 
    \For{$t=2:T_A$}
    \State Perform steps 17-24 and step 28 of Phase 2.
\EndFor
\State Phase 2: DDPG algorithm. 
\State Initialize replay buffer $\mathcal{B}$ and a random process $\alpha$.
    \For{$t=T_A,T_A+1,\ldots$}
    \State Select action $a_t=\mu(s_t|\theta_{\mu})+\alpha$.
    \State Scale action $a_t$ to satisfy the cache size constraint.
    \State Observe reward $r_t$ and new state $s_{t+1}$.
    \State Store sample $(s_t,a_t,r_t,s_{t+1})$ in the buffer $\mathcal{B}$.
    \State Get a random minibatch of $M_B$ samples $\{(s_i,a_i,r_i,s_{i+1})\}$ from $\mathcal{B}$.
    \State Set $y_i=r_i+\gamma Q'(s_{i+1},\mu'(s_{i+1}|\theta_{\mu}^{'})|\theta_Q^{'})$.
    \State Update the critic network by minimizing the loss: \\
    \hspace{1.5cm}$L(\theta_Q)=\frac{1}{M_B}\sum_i(y_i-Q(s_i,a_i|\theta_Q))^2$.
    \State Update the actor network using the sampled gradient according to (\ref{equ:a_gradient})
    \State Update the target networks\\
    \hspace{1.5cm}$\theta_{\mu}^{'}\leftarrow\tau\theta_{\mu}+(1-\tau)\theta_{\mu}^{'}$.\\
    \hspace{1.5cm}$\theta_Q^{'}\leftarrow\tau\theta_{Q}+(1-\tau)\theta_Q^{'}$.
\EndFor
\end{algorithmic}
\end{algorithm}


For the critic network, due to the different input types and dimensions, i.e., $s_t$ and $a_t$, we divide the neurons of the first hidden layer into two parts for feature extraction of $s_t$ and $a_t$ respectively.
\subsection{Supervised Learning for Pre-Training}
Like most model-free reinforcement learning algorithms, DDPG in general requires a large number of training episodes to find solutions. Moreover, the reward function in our considered problem involves both transmission delay and replacement cost and thus is difficult to be learned.
To tackle these issues, in this section, we use supervised learning to pre-train the actor and critic networks. The training samples are generated from the solution of a problem that minimizes the per-slot network cost instead of the total network cost by approximating the actual requests with the predicted ones. 

We first allocate the predicted number of requests $\widetilde{\mathbf{d}}(t)$ to each user based on the actual number of requests of each user in the previous time slot. Note that, it can also be allocated based on the number of requests from a previous window of time slots. For convenience, we only consider the previous time slot. The predicted transmission delay is calculated as:
\begin{equation}
\widetilde{C}_\text{d}(t)= \sum_{k\in\mathcal{K}}\sum_{f\in\mathcal{F}} \frac{d_{k,f}(t-1)}{\sum_{k' \in \mathcal{K}}d_{k',f}(t-1)}\widetilde{d}_f(t) D_k^{f}(t).
\end{equation}
Then we decouple the original problem (6) into a series of subproblems indexed by $t$. Each subproblem is to minimize the network cost at the current time slot and it is formulated as:
\begin{subequations}
\begin{align}
\mathcal{P}_A(t):&\min\limits_{\bm{\Lambda}(t)}\hspace{+0.2 cm}\widetilde{C}_\text{d}(t)+\beta C_\text{r}(t)\\
&\hspace{+0.08cm}\text{s.t.}\hspace{+0.31 cm} \sum_{f\in \mathcal{F}}\lambda_{f,n}(t)\leq M, \forall n\in\mathcal{N}.\\
&\hspace{+1.0 cm}\lambda_{f,n}(t)\in[0,1], \forall f\in\mathcal{F},n\in\mathcal{N}.
\end{align}
\end{subequations}

The problem $\mathcal{P}_A(t)$ is a convex problem which can be solved efficiently. Note that the parameters $[\mathbf{\widetilde{d}}(t),\bm{\Lambda}(t-1)]$ and the optimization variable $\bm{\Lambda}(t)$ of the problem $\mathcal{P}_A(t)$ correspond to the input $s_t$ and output $a_t$ of the actor network, respectively.
Hence, we use the parameters and the optimal variables of $\mathcal{P}_A(t)$ as the training set to pre-train the actor network. The $t$-th training sample $(x_t, y_t)$ of training set $\mathcal{T}_A$ is defined as $((\mathbf{\widetilde{d}}(t),\bm{\Lambda}(t-1)), \bm{\Lambda}^*(t))$,
where $\bm{\Lambda}^*(t)$ is the optimal solution of $\mathcal{P}_A(t)$.
The loss to minimize is
\begin{equation}
Loss(\theta_{\mu})=(y_t-\mu(x_t|\theta_\mu))^2. 
\end{equation}
After pre-training the actor network, we use the output of the actor network to pre-train the critic network. The method to pre-train the critic network is the same as the DDPG algorithm except that the actor network weights are not updated. We use the first $T_A$ time slots for pre-training.
After the pre-training phase, since the solution of $\mathcal{P}_A$ is only optimal to the approximate problem instead of the original problem, we use DDPG to further reduce the total network cost based on the pre-trained networks.

The proposed SDDPG algorithm is outlined in Alg.~\ref{SDDPG}.

\section{Simulations}
In this section, we evaluate the performance of the proposed request prediction algorithm and SDDPG algorithm by comparing them with several reference methods based on a real-world dataset.
\subsection{Dataset Description}
The dataset used in this paper is YouTube videos from Kaggle. It records an hourly real-time count observation (views, comments, likes, dislikes) during May 2018 of 1500 videos released in April 2018.
The total number of time slots is $T=600$. Since the number of requests in the dataset is recorded on a per-file basis without any user information, we allocate these requests uniformly at random to each user.
We select the 50 files with the largest total number of requests to be added to the catalogue, i.e., $F$ = 50.
\subsection{Simulation Setup}
\begin{itemize}
  \item \textbf{Wireless network setting}: We consider a hexagonal multi-cell network with $N=7$ cache nodes, each located at the center of a hexagonal-type cell. Each cache node can cache up to $M=5$ files. The length of each file $B$ is set to $1\text{GB}$. The distance between adjacent cache nodes is set to $500\text{m}$.
      There are total 20 users which are randomly and independently distributed in the network, excluding an inner circle of $50\text{m}$ around each cache node. The coverage area of cache node is $500\text{m}$. The transmission power of each cache node is set to $1\text{W}$. The available channel bandwidth for each user is set to $W = 0.1$MHz. The channel pathloss is modeled as $\text{PL(dB)}=148.1+37.6 \text{log}_{10}(d)$, where $d$ is the distance in kilometers.
      The transmit antenna power gain at each cache node is 1dBi.
      The noise power spectral density $\sigma^2$ for all users is set to $-152$dBm/Hz. The per-bit delay between each user and each cache node is calculated by $1/R$,  where $R=W\log_2(1+\text{SNR})$ with SNR being the average received signal-to-noise ratio.
      The MBS-to-user delay is thrice the maximum of the cache node-to-user delay.
      The discount factor $\gamma$ is set to 0.99.
  \item \textbf{C-LSTM setting}: For each cluster to do the prediction, we use a neural network with three hidden LSTM-layers, one input layer, and one output layer. The number of hidden units in each hidden layer is $24,24$ and $12$ respectively. We use a fully connected layer as the output layer and its activation function is set to be linear. The replay buffer size of each cluster is set to 1000. The learning rate and batch size are set to 0.0005 and 32 respectively.
  \item \textbf{Actor-Critic network setting}: We use two feed forward neural networks with one input layer, two hidden layers and one output layer to act as the actor and critic networks. The input of the critic network is the action $a_t$ and state $s_t$, and the output is the value of Q-function.
      The action $a_t$ and $s_t$ are assigned 200 neurons each in the first hidden layer of critic network. We add the outputs corresponding to the two parts of the neurons as the input of the second hidden layer, i.e., the input dimension of the second layer is 200. The number of neurons in the second layer is 100. Further, we use ReLU \cite{relu} as the activation function for the hidden layers. The activation function of output layer is set to be linear. The learning rate of critic network is set to 0.0005.
      A fully connected neural network is used as the actor network. The input of the actor network is the state $s_t$ and the output is the action $a_t$. The number of neurons in the first and second hidden layer are set to 800 and 400, respectively.
      ReLU is also used as the activation function for the hidden layers. The activation function of the output layer is sigmoid, which makes the output satisfy the cache fraction constraint.
      The learning rate of the actor network is set to 0.00001. The size of replay buffer $\mathcal{B}$ and the batch size $M_B$ are set to 1000 and 32 respectively. The trace parameter $\tau$ is set to 0.0005.
  \item \textbf{Pre-training setting}: In the pre-training phase, the pre-training time slot length $T_A$ is set to 500. CVX \cite{cvx} is used to solve the approximate problem.
      The learning rate and the batch size are set to 0.00005 and 32 respectively. ADAM \cite{adam} is chosen as the optimizer for all neural networks.
\end{itemize}

Note that the above hyperparameters are chosen based on experience and simulation results. Specifically, for the number of neurons in the hidden layer, we first set the number of neurons in the first hidden layer as the input dimension, and the number of neurons in each subsequent layer is half of the previous layer. Next, we increase or decrease the number of neurons in the hidden layer to find the optimal value while keeping other parameters unchanged.
For the learning rate, we start from 0.01 and reduce it by half each time to find the optimal value.

\subsection{Performance of Prediction}
We first evaluate the accuracy of the proposed prediction method by considering the normalized mean square error (NMSE). The NMSE at time slot $t$ is defined as follows:
\begin{equation}
E_N(t)=\frac{\|\widetilde{\mathbf{d}}(t)-\mathbf{d}(t)\|_2^2}{\|\mathbf{d}(t)\|_2^2}.
\end{equation}

We compare our method with the following benchmarks:

\begin{itemize}
  \item Grouped linear model (GLM) \cite{ZNF}: This method predicts the future requests by using a grouped linear regression model based on historical content requests, where the linear coefficients are designed in a grouped manner according to the age of each file.
  \item Long short-term memory (LSTM): Unlike C-LSTM, this scheme does not cluster features. Instead, each LSTM network is designed for one file and it only uses the features from the same file for prediction.
\end{itemize}

There are two key parameters in the proposed C-LSTM algorithm, the dimension of feature vector $\rho$ and the cluster number $C$. Increasing $\rho$ can introduce more information in the feature vector for prediction but also can increase the number of parameters of the LSTM network. Specifically, the number of parameters in the first LSTM layer is $4q(\rho+1+q)$ where $q$ is the number of neurons. As for the cluster number $C$, if it is too small,
the feature vectors of different trends can be classified into the same cluster, thus resulting in inaccurate prediction. If $C$ is more than needed, the feature vectors that have the same trends may be classified into different clusters. As a result, the number of samples in some clusters is small, which makes it difficult to train the LSTM network. In this subsection, we shall find the appropriate value for each parameter via simulation.

\begin{figure}[t]
\begin{centering}
\includegraphics[width=0.5\textwidth]{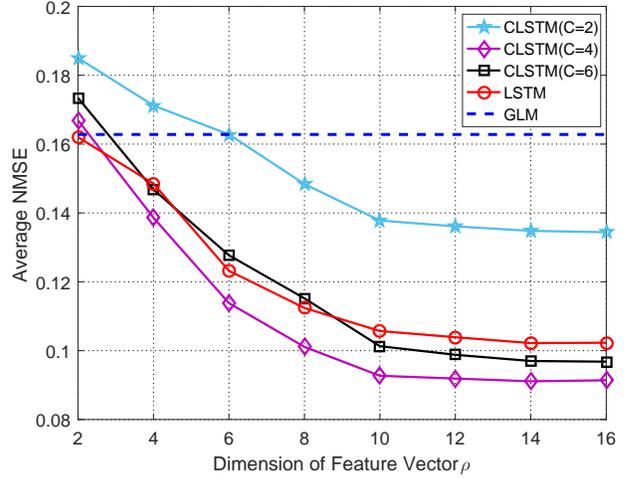}
 \caption{Average NMSE vs feature dimension.}\label{fig:window}
\end{centering}
\vspace{-0.3cm}
\end{figure}

Fig.~\ref{fig:window} plots the average NMSE with respect to the dimension of feature vector $\rho$. Since prediction is performed within a finite number of time slots, we compute the average NMSE by averaging the instantaneous NMSEs over all the time slots which is defined as $\bar{E}_N=\sum_{t=1}^T E_N(t)/T$.
From Fig.~\ref{fig:window}, we first observe that the average NMSE of LSTM and all C-LSTM under different $C$'s decreases and gradually approaches a constant as $\rho$ increases. This indicates that the prediction accuracy can be increased by increasing the historical observation window, but cannot be increased further if the window size is large enough. In particular, all the NMSE performances converge at around $\rho=12$. Therefore, we fix $\rho=12$, regardless of $C$, in the rest of our simulation to balance the average NMSE performance and the number of parameters to learn. From Fig. ~\ref{fig:window}, it is also observed that the
LSTM-based prediction methods can perform significantly better than the GLM method proposed in \cite{ZNF}. In particular,
when $\rho=12$ and $C=4$, the average NMSE of C-LSTM is $43.6\%$ lower than that of GLM. Furthermore, it is observed when $C=4$, the proposed C-LSTM is slightly worse than LSTM at small $\rho~(=2)$ , but becomes superior when $\rho\geq4$. In particular, the average NMSE of C-LSTM with $C=4$ is $11.6\%$ lower than that of LSTM at $\rho=12$. This is due to the fluctuation of feature vectors in low dimensions, which causes the feature vectors with different trends to be clustered together. As the dimension of feature vectors increases, the impact of fluctuations becomes small and the advantages of clustering gradually emerge. It is also seen from Fig.~\ref{fig:window} that the cluster size $C$ needs to carefully chosen, which shall be presented in the next figure.

\begin{figure}
\begin{centering}
\includegraphics[width=0.5\textwidth]{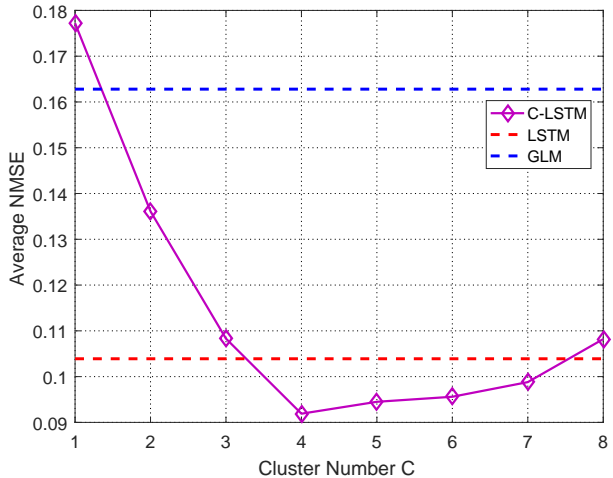}
 \caption{{Average NMSE vs cluster number.}}\label{fig:cluster number}
\end{centering}
\vspace{-0.3cm}
\end{figure}

Fig.~\ref{fig:cluster number} plots the average NMSE with respect to the cluster number $C$ when the dimension of feature vector $\rho=12$.
As expected, there exists an optimal number of clusters, which is $C=4$ in the considered example. If $C$ is not chosen properly, the prediction performance of C-LSTM can be worse than LSTM.
Specifically, when there is no clustering, i.e., $C=1$, C-LSTM is worse than GLM and LSTM. This is due to that the information contained in the feature vector is very different, and thus the average NMSE can be very large when the same LSTM network is used for prediction. When $C>1$, similar feature vectors are clustered together and each cluster is predicted by a separate LSTM network. By exploiting the correlation among the feature vectors in each cluster, the prediction performance can be enhanced. When $C$ is very large, the K-means algorithm becomes sensitive to the fluctuation in the feature vector, and feature vectors that should belong to the same cluster may be classified into different clusters. In addition, larger $C$ means fewer samples in each cluster, which makes it difficult to train the LSTM network. As a result, the average NMSE rises again when $C$ further increases.
Thus, we fix $C=4$ for the prediction part in the rest of the simulation.

\subsection{Performance of SDDPG}

We compare our method with the following benchmarks:

\begin{figure}[t]
\begin{centering}
\includegraphics[width=0.5\textwidth]{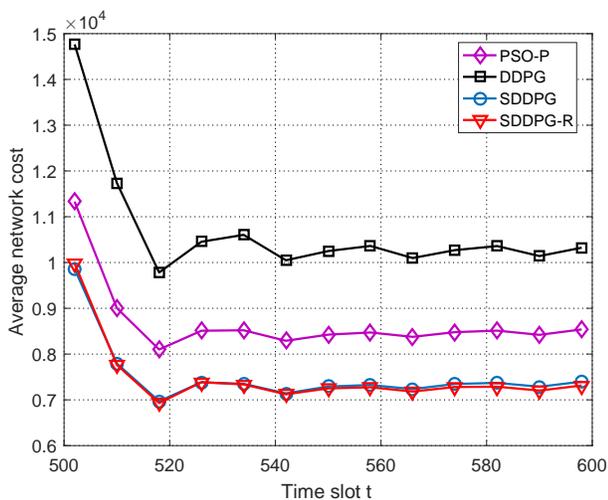}
 \caption{{Average network cost vs time slot.}}\label{fig:gamma1}
\end{centering}
\vspace{-0.2cm}
\end{figure}
\begin{itemize}
  \item Per-Slot Optimization with predicted requests (PSO-P): Use the predicted number of requests of the current time slot and the cache variable in the previous time slot as the input to solve the approximate problem $\mathcal{P}_A(t)$ directly. This method only minimizes the per-slot cost, but not the total cost.
  \item DDPG: This scheme differs from SDDPG in which there is no pre-training.
  \item SDDPG with real request (SDDPG-R): Replace the predicted number of requests in the state of SDDPG with the actual number of requests. This is an oracle scheme.
\end{itemize}

Fig.~\ref{fig:gamma1} plots the average network cost at $\beta=1.5$ with respect to time slot $t$ of the four schemes. The average network cost of time slot $t$ is defined by
$(\sum_i^t C(i))/t$. Please note that
the average network cost of all the four schemes shown in this
figure decreases rapidly during the time slot $[500, 520]$. This is due to the fact that the number of requests in our considered real-word dataset fall down after the peak time.
From Fig.~\ref{fig:gamma1}, we can observe that the average network cost of DDPG is much higher than that of the other three schemes. This means that it is difficult for neural network to find the optimal cache policy only based on the feedback from environment.
As such, the proposed supervised pre-training is essential to boosting the performance of DDPG. We also observe that the average network cost of SDDPG is $16.8\%$ lower than that of PSO-P and approaches the performance of SDDPG-R which is a lower bound.

\begin{figure}[t]
\begin{centering}
\includegraphics[width=0.5\textwidth]{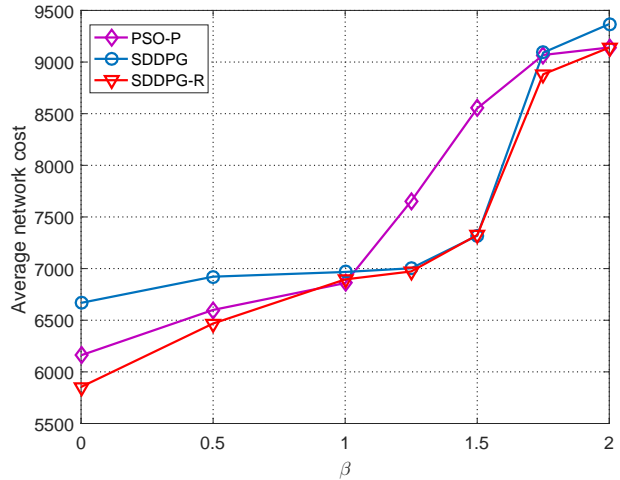}
 \caption{{Average network cost vs $\beta$.}}\label{fig:dif gamma}
\end{centering}
\vspace{-0.2cm}
\end{figure}
Fig.~\ref{fig:dif gamma} shows the average network cost with respect to weight $\beta$ that balances transmission delay and cache replacement cost. 
We can see that when $\beta=0$ (only transmission delay is considered), the performance of SDDPG is worse than that of PSO-P.
This is because when the replacement cost is not considered, minimizing the total network cost over all time slots is equivalent to minimizing the transmission delay at each time slot.
Obviously, the solution of PSO-P is already optimal. SDDPG, on the other hand, moves the solution away from the optimal solution by using stochastic gradient descent method. When $\beta>0$, we can find that the gap between SDDPG and PSO-P decreases first and then increases as $\beta$ becomes larger.  This is because the advantage of optimization over all time slots gradually becomes apparent as $\beta$ grows when $\beta$ is small. The gap between SDDPG and PSO-P reaches the maximum when $\beta=1.5$. When further increases, we can observe that the three schemes approach to perform the same.
This is because all three schemes tend to keep the cache state unchanged when $\beta$ is large enough.
Fig.~\ref{fig:dif gamma} also shows that the average network cost of the proposed SDDPG approaches that of SDDPG-R for a wide range of $\beta$, which again indicates that the prediction is accurate.
\subsection{Practical Coded Caching}
\begin{figure}[t]
\begin{centering}
\includegraphics[width=0.5\textwidth]{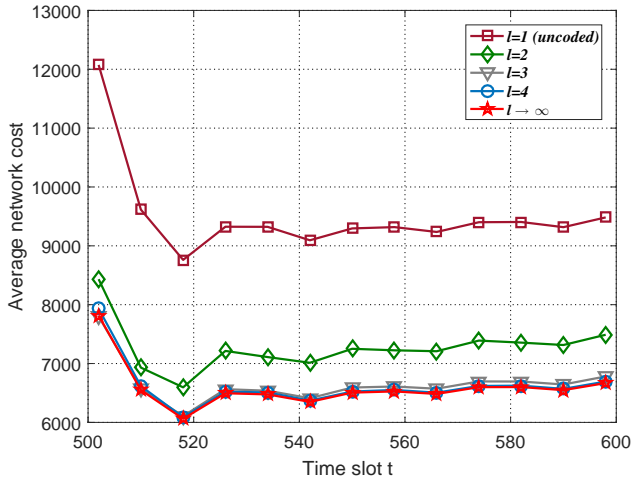}
 \caption{{Performance at different coding parameter $l$.}}\label{fig:dif_mds}
\end{centering}
\vspace{-0.2cm}
\end{figure}
In the previous simulation, we assumed an ideal coded caching scheme with a rateless MDS code so that the cache variable $\lambda_{f,n}(t)$ can take arbitrary value in $[0,1]$. In practice, considering the finite length of a file, we need to use a more practical coded caching scheme where each file can only be partitioned into a finite integer number of segments, denoted as $l\in\{1, 2, \ldots\}$ before MDS coding. That suggests that the cache variables ${\lambda_{f,n}(t)}$ can only take values from the discrete set $\{0, 1/l, 2/l, \ldots, (l-1)/l, 1\}$. When the element in the learned cache vector does not satisfy the discrete constraint, we choose $\widehat{\bm{\Lambda}}_n^*(t)$ which is the nearest neighboring cache vector within the alternative set
\begin{align} \mathcal{A}_n(t)\triangleq\{\widehat{\bm{\Lambda}}_n(t)|\widehat{\lambda}_{f,n}(t)\in &\{\lfloor \lambda_{f,n}(t)l \rfloor/l, \lceil \lambda_{f,n}(t)l \rceil/l\},\notag\\
&\sum_{f\in \mathcal{F}} \widehat{\lambda}_{f,n}(t)\leq M, f\in \mathcal{F}\}
\end{align}
as the approximate solution of $\bm{\Lambda}_n(t), \forall n\in \mathcal{N}$. More specifically:
{\begin{equation}
\widehat{\bm{\Lambda}}_n^*(t) = \arg\min\limits_{\widehat{\bm{\Lambda}}_n(t)\in\mathcal{A}_n(t)}
\|\widehat{\bm{\Lambda}}_n(t)-\bm{\Lambda}_n(t)\|_2.
\end{equation}}

Fig.~\ref{fig:dif_mds} shows the impact of the coding parameter $l$, where $l \rightarrow \infty$ and $l=1$ represent the ideal rateless MDS coding and no coding, respectively.  We can find that the performance increases as $l$ increases, since the cache decision can be better fine-tuned. It is observed that the performance with $l=4$ is very close to the case with $l\rightarrow \infty$. By comparing with $l=1$ (uncoded cache), it is also seen that coding brings significant cost reduction.


\section{CONCLUSION}
In this paper, we investigated the coded cache placement problem in a wireless network with multiple cache nodes using deep learning techniques. We formulated a cache placement problem of minimizing the total discounted network cost which involves transmission delay and replacement cost where the content popularity is unknown and dynamic. We first proposed C-LSTM approach to predict the number of requests using historical content requests. The correlation of the historical request information between different files was exploited to improve prediction accuracy. Based on the predicted result, we then proposed SDDPG approach that combines supervised learning and deep reinforcement learning to make cache decision. Real-world trace-based numerical results showed that the proposed C-LSTM approach can achieve higher prediction accuracy than the considered existing methods. The results also showed that the proposed SDDPG approach outperforms the per-slot optimization and DDPG without pre-training.
\bibliographystyle{IEEEtran}
\bibliography{IEEEabrv,iccc}

\begin{thebibliography}{10}
\providecommand{\url}[1]{#1}
\csname url@samestyle\endcsname
\providecommand{\newblock}{\relax}
\providecommand{\bibinfo}[2]{#2}
\providecommand{\BIBentrySTDinterwordspacing}{\spaceskip=0pt\relax}
\providecommand{\BIBentryALTinterwordstretchfactor}{4}
\providecommand{\BIBentryALTinterwordspacing}{\spaceskip=\fontdimen2\font plus
\BIBentryALTinterwordstretchfactor\fontdimen3\font minus
  \fontdimen4\font\relax}
\providecommand{\BIBforeignlanguage}[2]{{%
\expandafter\ifx\csname l@#1\endcsname\relax
\typeout{** WARNING: IEEEtran.bst: No hyphenation pattern has been}%
\typeout{** loaded for the language `#1'. Using the pattern for}%
\typeout{** the default language instead.}%
\else
\language=\csname l@#1\endcsname
\fi
#2}}
\providecommand{\BIBdecl}{\relax}
\BIBdecl

\bibitem{ICCC}
Z.~{Zhang} and M.~{Tao}, ``Accelerated deep reinforcement learning for wireless
  coded caching,'' in \emph{Proc. IEEE/CIC ICCC}, Aug 2019, pp. 249--254.

\bibitem{JSAC18}
G.~S. {Paschos}, G.~{Iosifidis}, M.~{Tao}, D.~{Towsley}, and G.~{Caire}, ``The
  role of caching in future communication systems and networks,'' \emph{IEEE J.
  Sel. Areas Commun.}, vol.~36, no.~6, pp. 1111--1125, June 2018.

\bibitem{femto}
K.~{Shanmugam}, N.~{Golrezaei}, A.~G. {Dimakis}, A.~F. {Molisch}, and
  G.~{Caire}, ``Femtocaching: Wireless content delivery through distributed
  caching helpers,'' \emph{IEEE Trans. Inf. Theory}, vol.~59, no.~12, pp.
  8402--8413, Dec. 2013.

\bibitem{zipfTCOM}
B.~N. {Bharath}, K.~G. {Nagananda}, and H.~V. {Poor}, ``A learning-based
  approach to caching in heterogenous small cell networks,'' \emph{IEEE Trans.
  Commun}, vol.~64, no.~4, pp. 1674--1686, Apr. 2016.

\bibitem{transfer2}
E.~{Baştuğ}, M.~{Bennis}, and M.~{Debbah}, ``A transfer learning approach for
  cache-enabled wireless networks,'' in \emph{Proc. IEEE WiOpt}, May 2015, pp.
  161--166.

\bibitem{Contcaching}
C.~{Zhong}, M.~C. {Gursoy}, and S.~{Velipasalar}, ``A deep reinforcement
  learning-based framework for content caching,'' in \emph{Proc. IEEE CISS},
  Mar. 2018, pp. 1--6.

\bibitem{living}
E.~{Bastug}, M.~{Bennis}, and M.~{Debbah}, ``Living on the edge: The role of
  proactive caching in 5g wireless networks,'' \emph{IEEE Commun. Mag.},
  vol.~52, no.~8, pp. 82--89, Aug. 2014.

\bibitem{ARAL}
J.~{Tang}, H.~{Tang}, N.~{Zhao}, K.~{Cumanan}, S.~{Zhang}, and Y.~{Zhou}, ``A
  reinforcement learning approach for {D2D}-assisted cache-enabled hetnets,''
  in \emph{Proc. IEEE GLOBECOM}, 2019, pp. 1--6.

\bibitem{letter}
P.~{Wu}, J.~{Li}, L.~{Shi}, M.~{Ding}, K.~{Cai}, and F.~{Yang}, ``Dynamic
  content update for wireless edge caching via deep reinforcement learning,''
  \emph{IEEE Commun. Lett.}, vol.~23, no.~10, pp. 1773--1777, 2019.

\bibitem{TVT}
G.~M.~S. {Rahman}, M.~{Peng}, S.~{Yan}, and T.~{Dang}, ``Learning based joint
  cache and power allocation in fog radio access networks,'' \emph{IEEE Trans.
  Veh. Technol.}, vol.~69, no.~4, pp. 4401--4411, 2020.

\bibitem{TWCMAB}
S.~{Müller}, O.~{Atan}, M.~{van der Schaar}, and A.~{Klein}, ``Context-aware
  proactive content caching with service differentiation in wireless
  networks,'' \emph{IEEE Trans. Wireless Commu}, vol.~16, no.~2, pp.
  1024--1036, Feb. 2017.

\bibitem{DynamicLocation}
P.~{Yang}, N.~{Zhang}, S.~{Zhang}, L.~{Yu}, J.~{Zhang}, and X.~{Shen},
  ``Dynamic mobile edge caching with location differentiation,'' in \emph{Proc.
  IEEE GLOBECOM}, Dec. 2017, pp. 1--6.

\bibitem{trendcache}
S.~{Li}, J.~{Xu}, M.~{van der Schaar}, and W.~{Li}, ``Trend-aware video caching
  through online learning,'' \emph{IEEE Trans. Multimedia}, vol.~18, no.~12,
  pp. 2503--2516, Dec. 2016.

\bibitem{DRLA}
L.~{Li}, Y.~{Xu}, J.~{Yin}, W.~{Liang}, X.~{Li}, W.~{Chen}, and Z.~{Han},
  ``Deep reinforcement learning approaches for content caching in cache-enabled
  {D2D} networks,'' \emph{IEEE Internet Things J.}, vol.~7, no.~1, pp.
  544--557, 2020.

\bibitem{ZNF}
N.~{Zhang}, K.~{Zheng}, and M.~{Tao}, ``Using grouped linear prediction and
  accelerated reinforcement learning for online content caching,'' in
  \emph{Proc. IEEE ICC Workshops}, May 2018, pp. 1--6.

\bibitem{LSTM11}
A.~Narayanan, S.~Verma, E.~Ramadan, P.~Babaie, and Z.-L. Zhang, ``Deepcache: A
  deep learning based framework for content caching,'' in \emph{Proc. of
  {NetAI}}, 2018, pp. 48--53.

\bibitem{LSTM15}
K.~C. {Tsai}, L.~{Wang}, and Z.~{Han}, ``Caching for mobile social networks
  with deep learning: Twitter analysis for 2016 u.s. election,'' \emph{IEEE
  Trans. Network Science and Engineering}, vol.~7, no.~1, pp. 193--204, 2020.

\bibitem{LSTM16}
Y.~Zeng and X.~Guo, ``Long short term memory based hardware prefetcher: a case
  study,'' in \emph{Proc. of ACM MEMSYS}, 2017, pp. 305--311.

\bibitem{lstm_gru}
N.~{Nguyen-Thanh}, D.~{Marinca}, K.~{Khawam}, S.~{Martin}, and L.~{Boukhatem},
  ``Multimedia content popularity: Learning and recommending a prediction
  method,'' in \emph{Proc. IEEE GLOBECOM}, Dec 2018, pp. 1--7.

\bibitem{JSAC2020}
V.~{Kirilin}, A.~{Sundarrajan}, S.~{Gorinsky}, and R.~K. {Sitaraman},
  ``Rl-cache: Learning-based cache admission for content delivery,'' \emph{IEEE
  J. Sel. Areas Commun.}, vol.~38, no.~10, pp. 2372--2385, 2020.

\bibitem{Xu}
X.~{Xu}, M.~{Tao}, and C.~{Shen}, ``Collaborative multi-agent multi-armed
  bandit learning for small-cell caching,'' \emph{IEEE Trans. Wireless Commu},
  vol.~19, no.~4, pp. 2570--2585, 2020.

\bibitem{pmab}
A.~{Sengupta}, S.~{Amuru}, R.~{Tandon}, R.~M. {Buehrer}, and T.~C. {Clancy},
  ``Learning distributed caching strategies in small cell networks,'' in
  \emph{Proc. IEEE. ISWCS}, Aug. 2014, pp. 917--921.

\bibitem{mugen}
Y.~{Zhou}, M.~{Peng}, S.~{Yan}, and Y.~{Sun}, ``Deep reinforcement learning
  based coded caching scheme in fog radio access networks,'' in \emph{Proc.
  IEEE ICC Workshops}, Aug. 2018, pp. 309--313.

\bibitem{sun}
H.~{Sun}, X.~{Chen}, Q.~{Shi}, M.~{Hong}, X.~{Fu}, and N.~D. {Sidiropoulos},
  ``Learning to optimize: Training deep neural networks for interference
  management,'' \emph{IEEE Trans. Signal Process}, vol.~66, no.~20, pp.
  5438--5453, Oct 2018.

\bibitem{weiyu}
W.~{Cui}, K.~{Shen}, and W.~{Yu}, ``Spatial deep learning for wireless
  scheduling,'' \emph{IEEE J. Sel. Areas Commun.}, vol.~37, no.~6, pp.
  1248--1261, June 2019.

\bibitem{yu16}
M.~{Eisen}, C.~{Zhang}, L.~F.~O. {Chamon}, D.~D. {Lee}, and A.~{Ribeiro},
  ``Learning optimal resource allocations in wireless systems,'' \emph{IEEE
  Trans. Signal Process}, vol.~67, no.~10, pp. 2775--2790, May 2019.

\bibitem{kmean}
J.~MacQueen \emph{et~al.}, ``Some methods for classification and analysis of
  multivariate observations,'' in \emph{Proc. 5th Berkeley Symp. Math. Statist.
  Probab}, vol.~1, no.~14.\hskip 1em plus 0.5em minus 0.4em\relax Oakland, CA,
  USA, 1967, pp. 281--297.

\bibitem{kmean++}
\BIBentryALTinterwordspacing
D.~Arthur and S.~Vassilvitskii, ``k-means++: The advantages of careful
  seeding,'' Stanford InfoLab, Technical Report 2006-13, June 2006. [Online].
  Available: \url{http://ilpubs.stanford.edu:8090/778/}
\BIBentrySTDinterwordspacing

\bibitem{LSTM1}
S.~Hochreiter and J.~Schmidhuber, ``Long short-term memory,'' \emph{Neural
  computation}, vol.~9, no.~8, pp. 1735--1780, 1997.

\bibitem{ls1}
A.~Graves and N.~Jaitly, ``Towards end-to-end speech recognition with recurrent
  neural networks,'' in \emph{Proc. ICML}, 2014, pp. 1764--1772.

\bibitem{ls2}
I.~Sutskever, O.~Vinyals, and Q.~V. Le, ``Sequence to sequence learning with
  neural networks,'' in \emph{Proc. NIPS}, 2014, pp. 3104--3112.

\bibitem{ls3}
O.~Vinyals, {\L}.~Kaiser, T.~Koo, S.~Petrov, I.~Sutskever, and G.~Hinton,
  ``Grammar as a foreign language,'' in \emph{Proc. NIPS}, 2015, pp.
  2773--2781.

\bibitem{ls4}
K.~Xu, J.~Ba, R.~Kiros, K.~Cho, A.~Courville, R.~Salakhudinov, R.~Zemel, and
  Y.~Bengio, ``Show, attend and tell: Neural image caption generation with
  visual attention,'' in \emph{Proc. ICML}, 2015, pp. 2048--2057.

\bibitem{DDPG}
T.~P. Lillicrap, J.~J. Hunt, A.~Pritzel, N.~Heess, T.~Erez, Y.~Tassa,
  D.~Silver, and D.~Wierstra, ``Continuous control with deep reinforcement
  learning,'' in \emph{Proc. ICLR}, 2016.

\bibitem{Dgradient}
D.~Silver, G.~Lever, N.~Heess, T.~Degris, D.~Wierstra, and M.~Riedmiller,
  ``Deterministic policy gradient algorithms,'' in \emph{Proc. ICML}, 2014, pp.
  387--395.

\bibitem{relu}
V.~Nair and G.~E. Hinton, ``Rectified linear units improve restricted boltzmann
  machines,'' in \emph{Proc. ICML}, 2010, pp. 807--814.

\bibitem{cvx}
M.~Grant and S.~Boyd, ``{CVX}: Matlab software for disciplined convex
  programming, version 2.1,'' 2014.

\bibitem{adam}
D.~P. Kingma and J.~Ba, ``Adam: A method for stochastic optimization,''
  \emph{arXiv preprint arXiv:1412.6980}, 2014.

\end{thebibliography}
\end{document}